\documentclass[useAMS,usenatbib]{mn2e} \usepackage{epsfig}
\usepackage{amsmath}
\usepackage{natbib}

\author[Kloska and Fortenberry] {Katherine A. Kloska$^{\dagger}$ and Ryan C.
Fortenberry$^{\ddagger}$\thanks{Email:rfortenberry@georgiasouthern.edu}  \\
$^\dagger$University of Kentucky, Department of Chemistry, Lexington, KY 40526,
U.S.A.\\
$^\ddagger$Georgia Southern University, Department of Chemistry \&
Biochemistry, Statesboro, GA 30460 U.S.A.}

\title[(MgO)$_n$]{Gas-Phase Spectra of MgO Molecules: A Possible Connection
from Gas-Phase Molecules to Planet Formation}

\begin{document}

\date{Submitted: \today}

\maketitle

\begin{abstract}

A more fine-tuned method for probing planet-forming regions, such as
protoplanetary discs, could be rovibrational molecular spectroscopy observation
of particular premineral molecules instead of more common but ultimately less
related volatile organic compounds.  Planets are created when grains aggregate,
but how molecules form grains is an ongoing topic of discussion in astrophysics
and planetary science. Using the spectroscopic data of molecules specifically
involved in mineral formation could help to map regions where planet formation
is believed to be occurring in order to examine the interplay between gas and
dust.  Four atoms are frequently associated with planetary formation: Fe, Si,
Mg, and O. Magnesium, in particular, has been shown to be in higher relative
abundance in planet-hosting stars.  Magnesium oxide crystals comprise the
mineral periclase making it the chemically simplest magnesium-bearing mineral
and a natural choice for analysis.  The monomer, dimer, and trimer forms of
(MgO)$_n$ with $n=1-3$ are analyzed in this work using high-level quantum
chemical computations known to produce accurate results.  Strong vibrational
transitions at 12.5 $\mu$m, 15.0 $\mu$m, and 16.5 $\mu$m are indicative of
magnesium oxide monomer, dimer, and trimer making these wavelengths of
particular interest for the observation of protoplanetary discs and even
potentially planet-forming regions around stars.  If such transitions are
observed in emission from the accretion discs or absorptions from stellar
spectra, the beginning stages of mineral and, subsequently, rocky body
formation could be indicated.


\end{abstract}

\section{Introduction}

The crossover in chemical classification from gaseous to solid matter is a
difficult question to answer in terms of astrophysics and astrochemistry.  The
low pressures and various temperature extremes of diverse astrophysical
environments often inhibit the creation of anything molecular beyond the gas
phase.  Yet, rocky bodies are comprised of solid matter.  How and when a
cluster of monomer molecules can begin to be classified as a mineral solid,
especially with regards to those that comprise bodies such as rocky planets, is
still not well understood \citep{Gail99, McWilliams12}.  The main hang-up for
this analysis is largely centered around what molecular species can be observed
as indicators of changes in the region being observed and mapped.

Most of the interstellar molecules detected thus far have been gaseous and have
atomic compositions from the upper-right of the periodic table
\citep{McCarthy01, Fortenberry17IJQC} since carbon, nitrogen, and oxygen are
among the most abundant elements in the universe \citep{Savage96}.  While many
of these have pertinence to planetary atmospheres, origins of life, and have
even been detected in protoplanetary discs \citep{Qi03}, few have been tied
directly to the formation of geologically-related solids and their constituent
minerals.  As such, reason exists to shift toward the other side of the
periodic table for searches related to rocky planets and how/where they might
form.  Most notably, [Al]/[Fe], [Si]/[Fe], [Sc]/[Fe], [Ti]/[Fe], and [Mg]/[Fe]
ratios have been found to be statistically-significantly higher in the spectra
of planet-hosting star systems \citep{Adibekyan12}.  Consequently, these atoms,
are likely important in small-mass planet formation, and the higher [Mg]/[Si]
ratio taken in the spectrum of a star or in that of a protoplanetary disc, the
more likely planets are to be found \citep{Adibekyan15}.  Since magnesium is
also one of the ten-most abundant atoms in the universe \citep{Savage96}
molecules containing magnesium should be fairly common.  As a result, magnesium
almost certainly plays a role in solid mineral and subsequent rocky body
formation, and molecules containing it could be utilized as indicators of such
processes.

Molecules containing one atom of magnesium have been previously detected in the
interstellar medium (ISM).  The carbon rich star IRC +10 216 has played host to
a significant portion of the known, unique interstellar molecules, and those
containing Mg are no exception.  Notably, these include: HMgNC detected in 2013
\citep{Cabezas13}, MgNC in 1993 \citep{Kawaguchi93}, MgCN in 1995
\citep{Ziurys95MgCN}, and MgCCH in 2014 \citep{Agundez14}.  While these forms
have been detected, there seems to be no direct tie between them and mineral
formation.

Alternatively, the spectrum of the as-of-yet undetected magnesium oxide (MgO)
clusters would be a natural place to begin such remote sensing.  In addition to
the abundance of magnesium, the presence of atomic oxygen in space is quite
prevalent, as it is the third most abundant element in the universe
\citep{Savage96}.  Although the formation of an Mg$-$O bond is unlikely in
hydrogen rich environments \citep{Kohler97}, the formation of an ionic bond
between oxygen and magnesium is inevitable due the atoms' abundances and large
electrostatic attraction. 

Furthermore, on Earth, magnesium oxide is the chemically simplest mineral
containing magnesium since it contains a repeating pattern of only two atoms.
In its cubic form, magnesium oxide creates the mineral periclase.  This mineral
occurs naturally in contact metamorphic rocks and also in the form of
ferropericlase (Fe,Mg)O in conjunction with iron making up about 20\% of the
Earth's mantle \citep{Ohta17}. Magnesium oxide has become well-known as a
refractory material and is remarkably resistant to changes when put under
extremely high temperatures and pressures \citep{Koker10}. In its solid form,
magnesium oxide is not magnetically conductive, but its superheated
liquid-phase is magnetic \citep{Coppari13}.  This liquid-phase may exist as
melt in the interiors of Super-Earths. Melt in terrestrial planet interiors can
enable magnetic fields of the planet due to their electrically conductive
characteristics \citep{Coppari13}.

In order for planetary formation to occur, all minerals must form from
molecules which must form from atoms in the gas-phase. These gas-phase
molecules aggregate eventually to form planetesimals which mark the beginning
of rocky planet formation \citep{Pollack96}. To determine where these gas-phase
molecules may be found and at what stage of planet formation they aggregate,
highly accurate spectroscopic data are needed for increasingly larger numbers
of MgO clusters.  Delineations in column densities or local abundances observed
in protoplanetary discs for the monomer would indicate gas phase chemistry,
while increased presence of the dimer, trimer, and larger clusters would give a
sign of a trend toward the solid and mineral phase, especially for spectral
features for each larger molecular cluster.  Such could be a marker for the
initiation of mineral or solid material formation, but the unique features of
each molecule must be available for reference.

Even though, the spectral features of magnesium oxide monomer were originally
observed by Herzberg over 50 years ago \citep{Herzberg66}, the dimer, trimer,
and higher $n$-mers have been little studied save for some anion
photodetachment analysis \citep{Gutowski00}.  The relative energies and
structural data for the gaseous clusters of (MgO)$_n$ ($n=1-40$), have been
previously calculated with density functional theory \citep{Chen08, Chen14,
Feitoza17} showing patterns of aggregation. Most notably, the face-centered
cubic patterns of (MgO)$_n$ begin to emerge at $n=4$ as a cube built from two,
stacked (MgO)$_2$ structures.  However, depending upon the $n$ value, cubic and
hexagonal isomers vie for the minimum energy structure \citep{Chen14}.  In any
case, spectral data for the smallest of these clusters will assist in the
characterization of magnesium oxide forms in the laboratory and potentially
even in mapping the physical properties of protoplanetary discs.


\section{Computational Details}

In the current work, the necessary vibrational frequencies and rotational
constants have been computed quantum chemically via a quartic force field, a
fourth-order Taylor series expansion of the internuclear Hamiltonian
\citep{Fortenberry13Morse} and is of the form:
\begin{equation}
V=\frac{1}{2}\sum_{ij} F_{ij} \Delta_i \Delta_j + \frac{1}{6}\sum_{ijk} F_{ijk}
\Delta_i \Delta_j \Delta_K + \frac{1}{24}\sum_{ijkl} F_{ijkl} \Delta_i \Delta_j
\Delta_k \Delta_l,
\label{QFF}
\end{equation}
in which the $F_{ij\dots}$ represents the force constants and the $\Delta_i$
terms describe the displacements.  In the past these types of calculations have
produced vibrational fundamental frequencies to within as good as  1.0
cm$^{-1}$ and rotational constants within 50 MHz of experimental data
\citep{Fortenberry11hoco, Fortenberry12HOCS+, Huang13NNOH+, Zhao14,
Fortenberry14C2H3+, Fortenberry14HOCS, Fortenberry15SiC2, Morgan15, Morgan152,
Fortenberry16C2H3+, Kitchens16, Bizzochi17}.  With these results, accurate
rovibrational molecular spectra can be produced, such that small clusters of
(MgO)$_n$ can be detected.

Coupled cluster theory \citep{ccreview, Shavitt09} at the singles, doubles, and
perturbative triples [CCSD(T)] level \citep{Rag89} as well as the explicitly
correlated second-order M{\o}ller-Plesset perturbation theory MP2-F12 levels
\citep{Rag89, MP2, Werner07, Hill10} are utilized to determine the spectroscopic
properties of the magnesium oxide clusters. The CCSD(T) method used to
calculate the spectroscopic data in this work is regarded as the ``gold
standard'' of quantum chemistry \citep{Helgaker04}.  The lower-level MP2
approach is significantly less costly than CCSD(T) which allows it to treat
larger systems, such as the magnesium oxide trimer here.  Additionally, MP2 is
known to produce surprisingly accurate results for a relatively low level of
theory due to the presence of a fortuitous cancellation of errors called a
``Pauling point'' in honor of Linus Pauling \citep{Zheng09, Sherrill09Rev,
Fink16}.  The MOLPRO 2015.1 program \citep{MOLPRO, MOLPRO-WIREs} as well as the
PSI4 program \citep{psi4, psi4JCTC} are used to perform the quantum chemical
computations.

\subsection{CCSD(T)-level Computations for MgO and (MgO)$_2$}
 

In order to form a QFF for the closed shell molecules, MgO and (MgO)$_2$, a
restricted Hartree-Fock CCSD(T)/aug-cc-pV5Z \citep{Dunning89, aug-cc-pVXZ,
cc-pVXZ, Prascher11} geometry optimization is performed to compute the initial
geometry.  The Martin-Taylor (MT) \citep{Martin94} core correlating basis set is
also used to modify the geometry to include core-orbitals, $1s$ for oxygen and
$1s2s2p$ for magnesium.  The reference geometries for these molecules are
determined by combining the CCSD(T)/aug-cc-pV5Z optimized geometry and the
differences in the CCSD(T)/MT optimized geometry with and without the core
orbitals.  This geometry optimization is performed in order to construct a
structure which is as close as possible to the true minimum.

From the reference geometries, the energy points for the anharmonic
internuclear  potential, which subsequently is used to produce the
spectroscopic data, are defined. The MgO monomer QFF requires 9 total points,
and the (MgO)$_2$ QFF requires 233 total points. The displacements of these
points, the $\Delta_i$ terms from Eq.~\ref{QFF}, are described as 0.005 \AA\ for
bond length coordinates and 0.005 radians for all bond angles and torsions.
The displacements are then used to define the Taylor series expansion and
refine the true minimum energy structure. 

\begin{figure}
\centering
\caption{The CcCR Optimized Geometry of Planar, $D_{2h}$ (MgO)$_2$.}
\includegraphics[width = 2.5 in]{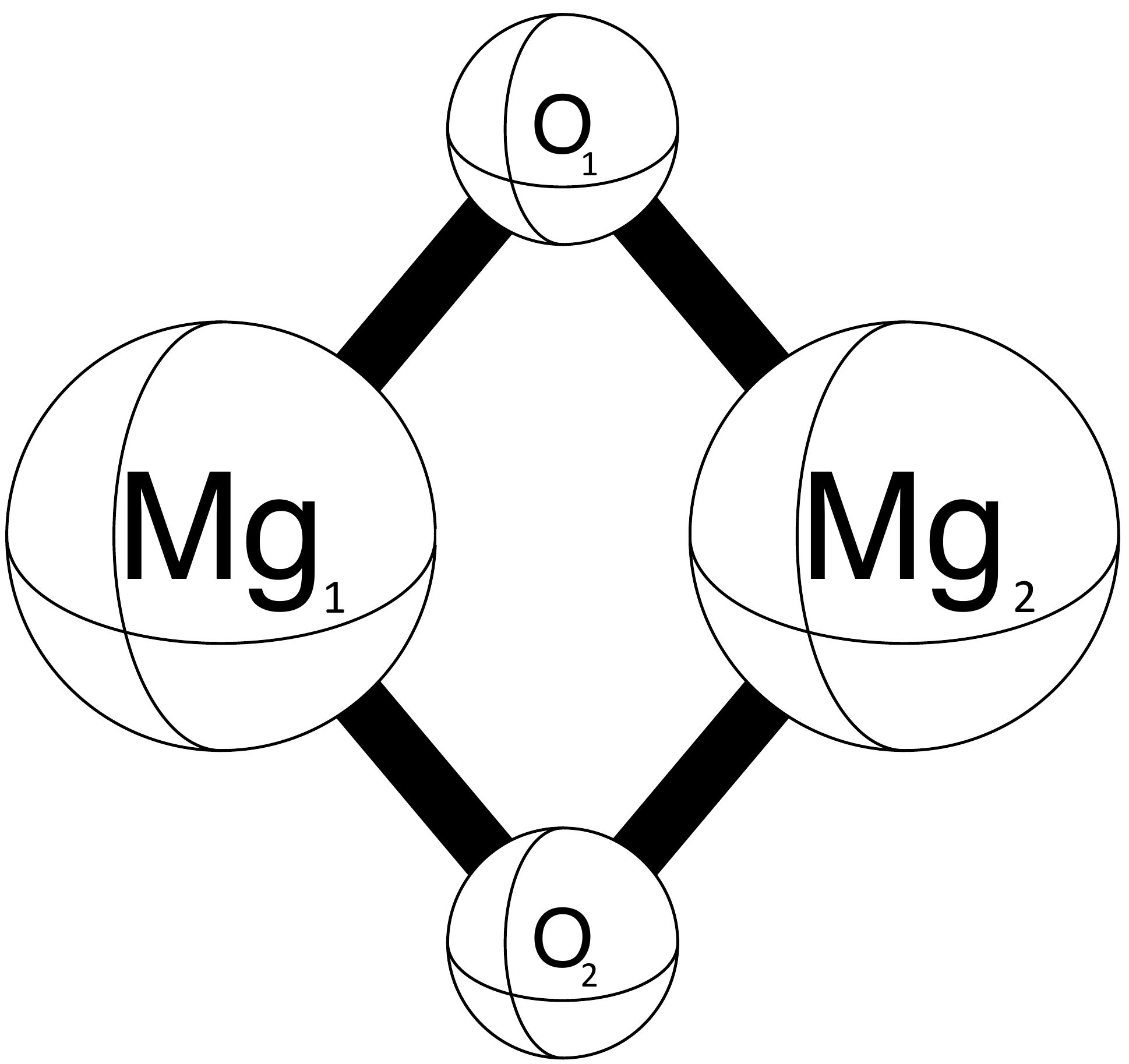}
\label{MgO2fig}
\end{figure}

Only one symmetry-internal coordinate defines the MgO molecule:
\begin{align}
S_{1}(\sigma) &= (\mathrm{O}-\mathrm{Mg}),
\end{align}
where the above coordinate is simply the distance between the oxygen and
magnesium atoms.  To fourth-order, there are four steps of positive
displacements, four steps of negative displacements, and the reference mimum
producing 9 total points.  The dimer, (MgO)$_2$, requires 233 total points
described by the following symmetry-internal coordinates:
\begin{align}
S_{1}(a_g) &= \frac{1}{\sqrt{2}}[(\mathrm{O}_1-\mathrm{O}_2)+(\mathrm{Mg}_1-\mathrm{Mg}_2)]\\
S_{2}(a_g) &= \frac{1}{\sqrt{2}}[(\mathrm{O}_1-\mathrm{O}_2)-(\mathrm{Mg}_1-\mathrm{Mg}_2)]\\
S_{3}(b_{1g}) &= \frac{1}{2}[(\mathrm{O}_1-\mathrm{Mg}_1)-(\mathrm{O}_1-\mathrm{Mg}_2)-(\mathrm{O}_2-\mathrm{Mg}_1)+(\mathrm{O}_2-\mathrm{Mg}_2)]\\
S_{4}(b_{1u}) &= \tau[(\mathrm{O}_1-\mathrm{Mg}_1-\mathrm{O}_2-\mathrm{Mg}_2)]\\
S_{5}(b_{3u}) &= \frac{1}{2}[(\mathrm{O}_1-\mathrm{Mg}_1)+(\mathrm{O}_1-\mathrm{Mg}_2)-(\mathrm{O}_2-\mathrm{Mg}_1)-(\mathrm{O}_2-\mathrm{Mg}_2)]\\
S_{6}(b_{2u}) &= \frac{1}{2}[(\mathrm{O}_1-\mathrm{Mg}_1)-(\mathrm{O}_1-\mathrm{Mg}_2)+(\mathrm{O}_2-\mathrm{Mg}_1)-(\mathrm{O}_2-\mathrm{Mg}_2)],
\end{align}
where the atom numbers are given in Figure \ref{MgO2fig}.

At each point, the energy produced is determined from CCSD(T). Using the
aug-cc-pVTZ, aug-cc-pVQZ, and aug-cc-pV5Z basis sets, the CCSD(T) energy can be
extrapolated using a three-point complete basis set (CBS) limit
\citep{Martin96}.  The energy differences between the CCSD(T) computations
retaining core orbitals and those lacking core orbitals again utilizing the MT
basis set are added to the CCSD(T)/CBS energies.   Additionally, the scalar
relativistic \citep{Douglas74} corrections are also added from computations
utilizing the cc-pVTZ-DK basis set for relativity included and excluded.  Using
these techniques to compute exceedingly accurate energies is defined as the CcCR
QFF \citep{Huang08, Huang09, Huang11, Fortenberry11hoco}. The ``C'' represents
the CBS energy, the ``cC'' represents the core correlation calculated as the
difference between the MT and MTc energies, and the ``R'' represents scalar
relativity. As mentioned above, these CcCR QFF computations are highly accurate
models that can physically describe spectroscopic data.

\subsection{MP2-F12 for (MgO)$_2$ and (MgO)$_3$}

\begin{figure}
\centering
\caption{The MP2-F12/aug-cc-pVDZ Optimized Geometry of Planar, $D_{3h}$ (MgO)$_3$.}
\includegraphics[width = 2.5 in]{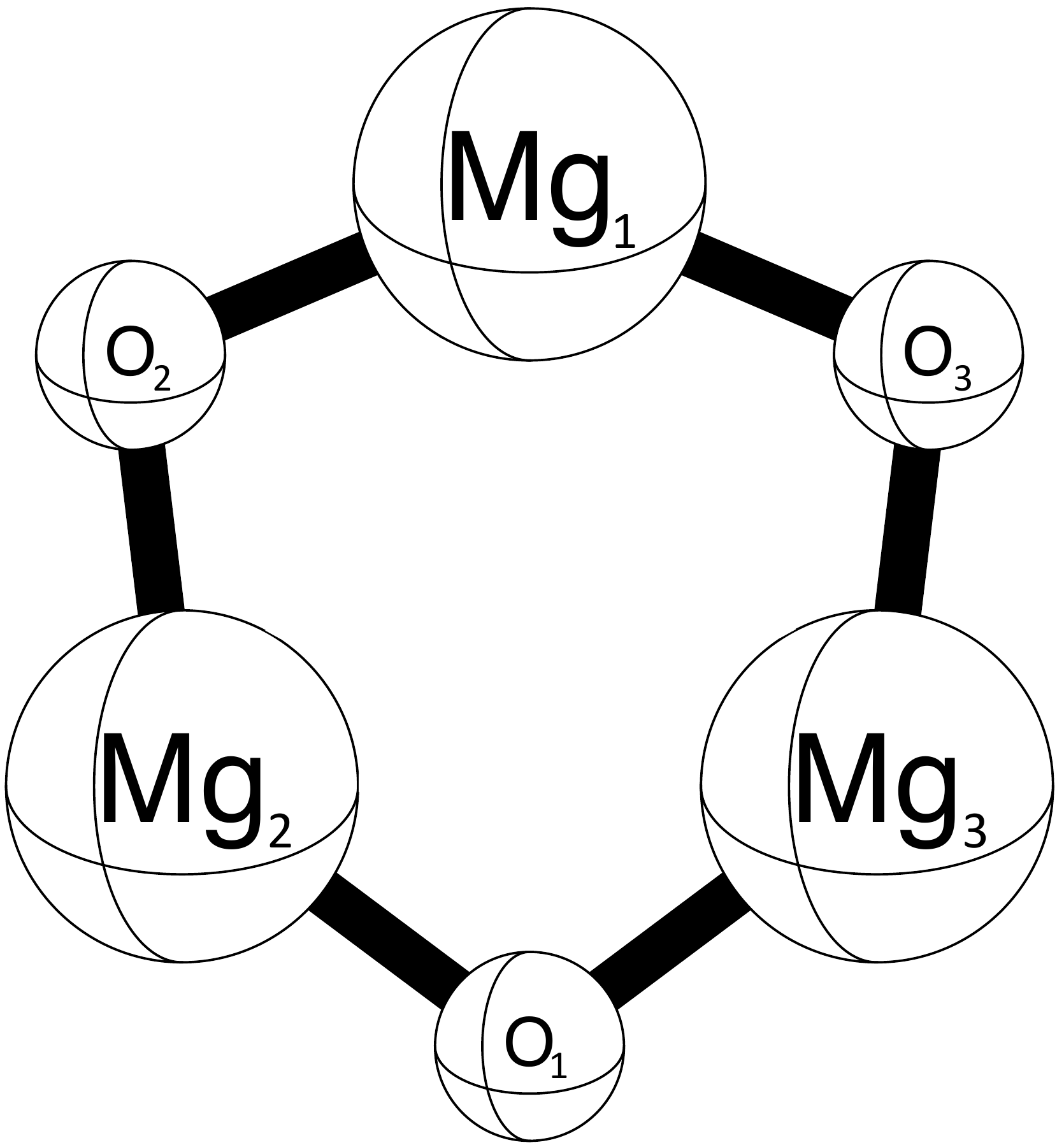}
\label{MgO3fig}
\end{figure}

The methodology for the MP2-F12 level has congruencies with the CCSD(T) level.
To begin, the initial geometry is optimized using the MP2-F12/aug-cc-pVDZ.
This optimized geometry is then used to define the necessary energy points to
form the anharmonic potential. Again, the (MgO)$_2$ QFF requires 233 total
energy points, but the (MgO)$_3$ QFF requires 3789 total energy points. The
displacements are still 0.005 \AA\ for bond length coordinates and 0.005
radians for all bond angles and torsions.  The coordinates for the dimer are
the same as those defined above for the CcCR QFF.  The trimer, (MgO)$_3$, is of
$D_{3h}$ symmetry, but will be run in $C_{2v}$ for ease in constructing the
points and in the force constants analysis.  The magnesium oxide trimer's QFF
is defined by 12 symmetry-internal coordinates:
\begin{align}
S_{1}(a_1) &= \frac{1}{\sqrt{2}}[(\mathrm{Mg}_1-\mathrm{O}_2)+(\mathrm{Mg}_1-\mathrm{O}_3)]\\
S_{2}(a_1) &= \frac{1}{\sqrt{2}}[(\mathrm{Mg}_1-\mathrm{Mg}_2)+(\mathrm{Mg}_1-\mathrm{Mg}_3)]\\
S_{3}(a_1) &= (\mathrm{Mg}_1-\mathrm{O}_1)\\
S_{4}(a_1) &= \frac{1}{\sqrt{2}}[(\mathrm{O}_2-\mathrm{Mg}_1-\mathrm{Mg}_2)+(\mathrm{O}_3-\mathrm{Mg}_1-\mathrm{Mg}_3)]\\
S_{5}(a_1) &= \frac{1}{\sqrt{2}}[(\mathrm{O}_2-\mathrm{Mg}_1-\mathrm{O}_1)+(\mathrm{O}_3-\mathrm{Mg}_1-\mathrm{O}_1)]\\
S_{6}(b_2) &= \frac{1}{\sqrt{2}}[(\mathrm{Mg}_1-\mathrm{O}_2)-(\mathrm{Mg}_1-\mathrm{O}_3)]\\
S_{7}(b_2) &= \frac{1}{\sqrt{2}}[(\mathrm{Mg}_1-\mathrm{Mg}_2)-(\mathrm{Mg}_1-\mathrm{Mg}_3)]\\
S_{8}(b_2) &=\frac{1}{\sqrt{2}}[\angle(\mathrm{O}_2-\mathrm{Mg}_1-\mathrm{Mg}_2)-\angle(\mathrm{O}_3-\mathrm{Mg}_1-\mathrm{Mg}_3)]\\
S_{9}(b_2) &=\frac{1}{\sqrt{2}}[(\mathrm{O}_2-\mathrm{Mg}_1-\mathrm{O}_1)-(\mathrm{O}_3-\mathrm{Mg}_1-\mathrm{O}_1)]\\
S_{10/x}(b_1) &= \frac{1}{\sqrt{2}}\tau[(\mathrm{O}_3-\mathrm{Mg}_1-\mathrm{Mg}_3-\mathrm{O}_1)-(\mathrm{O}_2-\mathrm{Mg}_1-\mathrm{Mg}_2-\mathrm{O}_1)]\\
S_{11/y}(b_1) &= \tau(\mathrm{O}_2-\mathrm{Mg}_1-\mathrm{O}_3-\mathrm{O}_1)\\
S_{12/z}(a_2) &= \frac{1}{\sqrt{2}}\tau[(\mathrm{O}_3-\mathrm{Mg}_1-\mathrm{Mg}_3-\mathrm{O}_1)+(\mathrm{O}_2-\mathrm{Mg}_1-\mathrm{Mg}_2-\mathrm{O}_1)].
\end{align}
At each point, MP2-F12/aug-cc-pVDZ energies are computed, and this QFF is only
defined by these terms.  The atom labels are shown in Figure \ref{MgO3fig}.

\subsection{Forming the QFF with CCSD(T) and MP2-F12 methods}

After obtaining the set of energies, the QFF is fitted. A least-squares fitting
of the energy points (whether CcCR or MP2-F12) produces results with minimal
error as indicated by the sum of squared residuals of less than 10$^{-16}$
a.u.$^2$ (10$^{-18}$ for the monomer and dimer) for all levels of theory used
and molecules examined. This fitting produces the equilibrium geometry.
Refitting once more zeroes the gradients and produces the force constants
needed to define the potential portion for the internuclear Hamiltonian, the
QFF (Eq.~\ref{QFF}).

The INTDER \citep{intder} program transforms the computed force constants into
Cartesian coordinates which are more readily translated for subsequent generic
analysis.  Rotational and vibrational second-order perturbation theory (VPT2)
\citep{Mills72, Watson77, Papousek82} are employed by combining the previously
computed potential energy and the new kinetic energy via the
SPECTRO \citep{spectro91} program.  These analyses provide the anharmonic
vibrational frequencies and spectroscopic data.   (MgO)$_2$ possesses a
$2\nu_6=\nu_1$ type-1 Fermi resonance as well as $\nu_3/\nu_2$ and
$\nu_5/\nu_4$ C$-$type Coriolis resonances.  (MgO)$_3$ exhibits dozens of
type-1 and type-2 Fermi, Coriolis, and Darling-Denison resonances, but none
perturb the vibrational frequencies by more than 2 cm$^{-1}$ which is well
within the expected accuracy of MP2-F12/aug-cc-pVDZ approach utilized here.
All vibrational intensities are computed within the double-harmonic
approximation and with MP2/6-31+G$^*$ in the Gaussian09 program \citep{MP2, g09,
Hehre72}.  This approximation has been shown to be in good agreement with
higher-level computations for the prediciton of such properties
\citep{Fortenberry14LLL, Yu15NNHNN+}.

\section{Results and Discussion}

\subsection{MgO Singlet and Triplet}

\begingroup
\begin{table}

\caption{The MgO CcCR Force Constants in mdyn/\AA$^n$ for $n,m=0,1,2,3,4$.}

\label{MgOfc}

\centering

\begin{tabular}{c r | c r c r}
\hline
\multicolumn{2}{c|}{$\tilde{X}\ ^1\Sigma^+$ MgO} & \multicolumn{2}{c}{$A\ ^3\Pi$ MgO} \\
\hline
F$_{11}$ &  3.763 052 & F$_{11}$ &  2.384 934  \\ 
F$_{111}$ & -22.118 502 & F$_{111}$ & -11.858 582  \\
F$_{1111}$ &  66.966 674 & F$_{1111}$ & 52.025 584  \\

\hline

\end{tabular}

\end{table}
\endgroup

\begingroup
\begin{table*}
	
\caption{The MgO Singlet and MgO Triplet CcCR QFF VPT2 Bond Lengths, Spectroscopic Constants, and Vibrational Frequencies (with intensities in km/mol in parentheses)}

\label{MgO}

\centering

\begin{tabular}{l r|r r r r r |r r}
\hline
&& \multicolumn{5}{c|}{$\tilde{X}\ ^1\Sigma^+$ MgO} & \multicolumn{2}{c}{$A\ ^3\Pi$ MgO} \\
&& MgO & $^{25}$MgO & $^{26}$MgO & Mass Avg. & Experiment$^a$ & Mass Avg. & Experiment$^b$\\
\hline
$r_e$ (O-Mg) & \AA       & 1.738 981 & --        & --        & --        & 1.7490    &1.876 550 & 1.869 286\\
$B_e$        & MHz       & 17416.0   & 17136.9   & 16880.3   & 17329.2   & --        &14881.5   & -- \\ 
Est.~$B_0$   & MHz       &           &           &           & 17242.5   & 17233.31  &14807.1   & 15072.5 \\ 
$D_e$        & kHz       & 35.331    & 34.208    & 33.191    & 35.005    & 36.937    &34.9574   & 35.92\\
$H_e$        & mHz       &  -29.838  &  -28.426  &  -27.268  &  -29.440  &--         &-8.933    & -- \\
\hline
$\omega_1$   & cm$^{-1}$ & 814.8 (1127) & 809.3 & 803.2 & 813.0 & 785.2 & 647.9 (61) & 650.223 \\
\hline
\end{tabular}
\\$^a$ \cite{Irikura07} with $D$ from \cite{Murtz95}.
\\$^b$ \cite{Murtz95}.
	
\end{table*}
\endgroup

In order to determine accurate rovibrational spectroscopic data for (MgO)$_n$
as a potential marker of mineral formation in protoplanetary discs, the
analysis begins with the monomer, MgO. The CcCR symmetry-internal harmonic
force constants for $\tilde{X}\ ^1\Sigma^+$ MgO and $A\ ^3\Pi$ MgO are listed
in Table \ref{MgOfc}, and define the QFF.  The singlet is more tightly bound in
the F$_{11}$ value indicating a higher bond order.  Magnesium is less inclined
to form double bonds because this atom has a [Ne]$3s^2$ electron configuration.
Sharing or outright removal of those $s$ orbitals are much more likely than any
notable $p$ orbital population, which is  a characteristic of a double bond.
In fact, the triplet has a strength of merely 63\% to that of the singlet.  The
implication is that $\tilde{X}\ ^1\Sigma^+$ MgO is more ionic in character
since one of the electrons from the magnesium atom now occupies a lone pair
orbital on the carbon in order to create the spin-pairing of the singlet.  This
relationship is somewhat unexpected since it is opposite of that noted for the
isoelectronic MgCH$_2$ molecule recently analyzed in our group where the
triplet state is the ground state \citep{Bare00, Bassett17MgCH2}.  However,
this has been known in MgO for over 40 years \citep{Ikeda77} and the more
contemporary 2551.9713 cm$^{-1}$ $a ^3\Pi \leftarrow \tilde{X} ^1\Sigma^+$
transition frequency \citep{Murtz94, Murtz95} is very close to the 2385
cm$^{-1}$ CcCR adiabatic excitation computed here.  Regardless, MgO monomer
will almost exclusively exist in the $\tilde{X}\ ^1\Sigma^+$ state in
protoplanetary discs at distances where rocky planets will form since the
excitation energy is equivalent to an ambient temperature of roughly 3700 K.
 
The CcCR QFF VPT2 bond lengths and spectroscopic constants are displayed in
Table \ref{MgO}.  The O$-$Mg bond is longer in the triplet state than in the
singlet state as a result of the smaller F$_{11}$ value in the triplet.  While
$\tilde{X}\ ^1\Sigma^+$ MgO is more ionic than $A\ ^3\Pi$ MgO, the monomer can
be more tightly bound since both electrons can be involved in the bonding.
However, in the triplet, the electron is reatined by the magnesium but lies on
the side opposite the carbon.  Therefore, it is unavilable for bonding weaking
the Mg$-$O interaction.  The $\tilde{X}\ ^1\Sigma^+$ MgO bond length is
comparable to the experimental bond length of 1.7490 \AA\ \citep{Irikura07},
and even the excited triplet state is within 0.02 \AA\ of experiment at 1.869
286 \AA\ \citep{Murtz95}.  The variance in the bond lengths, rotational
constants, and sextic distortion constants between the singlet and triplet
states showcases that these two forms of MgO will be easily distinguished
rotationally as well as electronically or by ambient temperature.

However, the computed rotational constant for $\tilde{X}\ ^1\Sigma^+$ MgO does
not match that from experiment 17233.31 MHz \citep{Irikura07}.  Magnesium has
three major isotopes, unlike oxygen, carbon, and hydrogen where one dominates.
$^{24}$Mg has a natural abundance of 79\%, $^{25}$Mg is 10\%, and $^{26}$Mg is
11\%.  By averaging the equilibrium rotational constants, $B_e$, for each
isotopologue, 17329.2 MHz is produced, in much better agreement.  Full
vibrational-averaging for diatomics is problematic for the available software.
However, estimates in the perturbation between $B_e$ and the more
experimentally-meaningful $B_0$ values from data taken below for (MgO)$_2$ give
a rotational constant within 10 MHz of experiment at 17242.5 MHz.  Hence, our
experimental methods should be similarly accurate for the dimer species below.
Case in point are the quartic distortion contants ($D$) where is computed to be
35.004 kHz and experiment which is 36.937 kHz \citep{Murtz95}.  The $A\ ^3\Pi$
bond length, rotational constant, $D$ value, and even vibrational frequecny are
in excellent agreement with experiment \citep{Murtz95} showcasing the accuracy
of this computational approach.
 
The vibrational frequencies of MgO are also listed in Table \ref{MgO}. The
singlet state has an incredibly bright intensity (1127 km/mol) for the
fundamental mode due to an active charge transfer which is a hallmark of ionic
species.  Again, anharmonic computations are problematic for diatomics, but
mass-averaging lowers the $\omega_1$ fundamental frequency somewhat to 813.0
cm$^{-1}$.  Considering an anharmonic effect on the order of 10-20 cm$^{-1}$
puts the estimated frequency within 10 cm$^{-1}$ of the experimentally-observed
value at 785.2 cm$^{-1}$.  Consequently, any computational studies of magnesium
must involve all three major isotopes in order for meaningful comparison to be
had for physically observable properties whether in planet-forming regions or
in the laboratory.  

\subsection{(MgO)$_2$}

The dimer, (MgO)$_2$, requires significantly more force constants than the
monomer. These force constants are produced in Table 1 of the Supporting
Information (SI).  The higher $D_{2h}$ symmetry means that the bond strengths
arise from linear combinations of the force constants.  As a result, the
bonding in (MgO)$_2$ is somewhat weaker between individual Mg$-$O bonds, but
the cross-bonding between like atoms in the ring and the cyclic stabilization
create a more stable structure once two MgO pieces are brought together.  In
fact, the CcCR separation energy for (MgO)$_2$ $\rightarrow$ 2 MgO is -132.0
kcal/mol indicating a strong stabilization for creation of the dimer from the
monomer.  Granted some barrier will likely be present in any such reaction, but
the thermodynamics will ultimately lead to the larger clusters.  Each MgO unit
contains 20 electrons meaning that each successive $n$-mer will be add a factor
of 20 electrons to the system making for simple addition.

The vibrational frequencies of (MgO)$_2$ are shown in Table \ref{MgO2vib}.  The
anharmonic corrections are well-behaved within our tight fitting giving no
reason to doubt the accuracy of these values as within the 10 cm$^{-1}$ or less
range established for MgO above and previously for $p$-block molecules.  The
mass averaging is of supreme importance here for observations of molecules
mixed through a bulk.  While the ($^{24}$MgO)$_2$ isotopologue will make up
62\% of the observable amount of (MgO)$_2$, $^{24}$MgO$^{25}$MgO and
$^{24}$MgO$^{26}$MgO will comprise 7.9\% and 8.7\% of (MgO)$_2$.  Hence, these
and the three other combinations of magnesium masses are included in the
averaged values given in the furthest column of Table \ref{MgO2vib}.

Only three of the six fundamental vibrational frequencies are infrared-active
due to symmetry, $\nu_2$ (the oxygen atom translation), $\nu_3$ (the magnesium
atom translation), and $\nu_6$ (the out-of-plane ``book'' motion) as given in
Table \ref{MgO2vib}.  These three are all notably bright frequencies with
intensities above 100 km/mol.  Most molecules have intensities around a few
dozen km/mol.  As a result, the magnesium oxide dimer should be detectable even
in reduced amounts as a consequence of Beer's law.  The 15-17 $\mu$m region at
the beginning of the far-IR and approaching the THz region is well within the
range of the James Webb Space Telescope (JWST) making this upcoming observatory
a natural choice for future observation of the $\nu_2$ fundamental vibrational
frequency of (MgO)$_2$.  The 281.4 cm$^{-1}$ (35.5 $\mu$m or 8.44 THz)
out-of-plane motion will be below what JWST can observe, but this is the least
bright fundamental of the three.  The frequencies for the two-quanta overtones
for each of these fundamentals are also given in Table \ref{MgO2vib}.

The differences between the CcCR and MP2-F12/aug-cc-pVDZ calculations differ by
3-5\%.  Hence, MP2-F12/aug-cc-pVDZ is well-behaved, but undershoots the
prediction of the frequencies by as much as 30 cm$^{-1}$.  However, this is
close enough use in conjunction with isolated laboratory experiments for larger
molecules where such anharmonic computations cannot be undertaken with the CcCR
QFF.

Even though, (MgO)$_2$ is non-polar, the CcCR QFF structures and spectroscopic
constants are included in Table \ref{MgO2geom} for completeness.  The
vibrationally-averaged ($R_{\alpha}$) bond lengths for (MgO)$_2$ differ
slightly between the CcCR and MP2-F12/aug-cc-pVDZ methods as expected for
different methods and basis sets.  However, the spectroscopic $D$ and $H$
constants are consistent between the two approaches with the primary rotational
constants roughly similar.  The near-oblate nature of this molecule means that
the $A$ and $B$ constants are closer to one another in magnitude and greater
than the $C$.  Even without a permanent dipole moment, rotational transitions
of vibrationally-excited states of molecules have been detected in the
ISM \citep{Turner87, Cernicharo08} making the present predictions for the $A$,
$B$, and $C$ constants of the 2, 3, and 6 states potentially of value for
observations with the Atacama Large Millimeter Array (ALMA).
 
 
\subsection{(MgO)$_3$}

Creating the magnesium oxide trimer from three monomers will also produce 268.0
kcal/mol indicating that the trimer is even more stable than the dimer since
half of this stabilization energy is 134.0 kcal/mol, 2.0 kcal/mol more than the
dimer.  Even though the two are computed with different methods
(MP2-F12/aug-cc-pVDZ versus CcCR), the relative energies are still directly
comparable.  The slightly larger harmonic, diagonal (MgO)$_3$ force constants,
displayed in Table 2 of the SI, also corroborate the growth in stabilization
for larger clusters of (MgO)$_n$. Since periclase or bulk magnesium oxide is a
stable ionic lattice, such an increase in stabilization for larger clusters is
fully expected.  As a result, the larger clusters are behaving more like the
mineral with each addition of a monomer unit \citep{Chen14}.

While there are 10 possible combinations of magnesium isotopes possible in
(MgO)$_3$, only the six that contain at least one $^{24}$Mg will have natural
abundances of greater than 0.4\%.  Hence, the isotopes included in Tables
\ref{MgO3vib} and \ref{MgO3geom} will make up more than 98\% of any observable
(MgO)$_3$.  Only these are included in the mass averaging.

Of the twelve total fundamental vibrational frequencies, four are in degenerate
pairs making a total of eight fundamental frequencies.  Even though the QFF is
computed in the lower-symmetry $C_{2v}$ point group, the vibrational
frequencies are computed in full $D_{3h}$.  Of the eight fundamental
frequencies, four are vibrationally-active, but the $\nu_3$ motion is nearly
unobservable.  As a result and like with (MgO)$_2$, the magnesium oxide trimer
has three observable fundamental vibrational frequencies (with atom labels from
Fig.~\ref{MgO3fig}): $\nu_2(E')$, the Mg$_2$+Mg$_3$/O$_2$+O$_3$ stretch akin to
the vibrationally-active mode in H$_3$$^+$; $\nu_6(A_2'')$, the Mg triangle and
O triangle out-of-plane separation; and $\nu_7(E')$, the Mg$_1$+O$_1$ stretch).

The anharmonic shifts are small for the first four fundamentals.  After that,
the anharmonicities are fairly large and actually increase the fundamental
frequency.  Such behavior is not typical but not uncommon.  These modes are
low-frequency in the same range where most positive anharmonicities are
reported in other highly-symmetric systems.  However, $\nu_5-\nu_8$ should
still be viewed as less trustworthy than the typically-behaving fundamentals in
$\nu_1$-$\nu_4$.  Follow-up experiments should be able to determine the $\nu_6$
and $\nu_7$ peak positions due to the large intensities of these motions.
Regardless, these frequencies are not observable with JWST anyway.

The only observable fundamental frequency for (MgO)$_3$ in JWST's range is
$\nu_7$ which VPT2 with the MP2-F12/aug-cc-pVDZ QFF reports the mass averaged
fundamental to be at 748.6 cm$^{-1}$.  Taking the benchmark for the performance
of the dimer and upscaling this value by 5\% to perform more like the CcCR QFF
puts the fundamental for this bright transition at roughly 785 cm$^{-1}$,
12.7$\mu$m, or 23.5 THz.  This is in the exact range as the MgO monomer.
Hence, their vibrational spectra will likely rest at nearly the same frequency.

According to Table \ref{MgO3geom} the mass averaged bond length in this
molecule is 1.848 \AA, shorter than the 1.860 \AA\ mass averaged bond length in
(MgO)$_2$ giving further indication of stronger bonding as discussed
previously.  This bond length does not vary significantly as the molecule
includes the $^{25}$Mg and $^{26}$Mg isotopes such that the standard isotopologue
and the other masses have the same structure for all intents and purposes.  As
with the dimer, (MgO)$_3$ has no dipole moment and will not be
rotationally-observable.  However, the rotational constants and spectroscopic
data are provided for completeness.  Magnesium oxide trimer is even more
nearly-oblate than dimer as implied by the rotational constants and as expected
by the structure.  Again, the rotational spectra of some of the
vibrationally-excited states could be observed.  The accuracies for such data
with MP2-F12/aug-cc-pVDZ would not be as good as desired, but the pure ground
vibrational state rotational data should be a good first-order approximation
for their values.
 

\section{Conclusions} 

Since magnesium is known to be in higher relative abundances in stars that host
planets and magnesium is believed to comprise a significant portion of
terrestrial-type planets, molecules containing element-12 could serve as
sentinel markers in planet forming regions where gas phase-chemistry is giving
way to the creation of solids, minerals most notably.  The simplest magnesium
mineral, magnesium oxide or periclase, is a good place to begin such analysis.
The advent of JWST will bring high-resolution to the observation of
protoplanetary discs and planet forming regions around stars. This work
provides data for the detection of the (MgO)$_n$ molecular clusters for
$n=1-3$.  Such molecules are the building blocks of larger magnesium oxide
crystals.  Hence, if the $n\geq2$ molecular clusters can be detected in the
gas-phase in such planet-forming environments as protoplanetary discs, they
will be clear indicators of larger mineral formation moving from the gas- to
solid- (mineral) phases.

This work has shown that magnesium oxide clusters are more strongly bound as
monomer units are added.  Such is expected for premineral molecules that
ultimately build larger clusters and eventually ionic lattices in solid
materials.  Additionally, the excellent monomer CcCR QFF benchmarks show that
the equivalently computed dimer spectral features should be equivalently
accurate.  Therefore, scaling-up the trimer MP2-F12/aug-cc-pVDZ results will
provide similar results, especially for the higher-frequency fundamentals.

In general, transitions at 12.5 $\mu$m (monomer \& trimer), 15.0 $\mu$m
(dimer), and 16.5 $\mu$m (dimer) will be clear indicators of initial periclase
formation from these MgO clusters.  These transitions also fall within a
relatively clear window of observed spectra for known protoplanetary discs
\citep{Williams11}.  They are to the red of a dominant silicate feature at 10
$\mu$m making these peaks observable as shoulders or simply separate,
lower-frequency peaks especially if the resolving power of JWST can be brought
to bear.  Since the 785 cm$^{-1}$/12.5 $\mu$m band is exhibited for both the
monomer and the trimer, the tetramer will likely have a transition very close
to this range.  As a result, mapping the 12.5 $\mu$m band in protoplanetary
discs while also mapping the growth of the dimer's 16.5 $\mu$m band and/or the
reduction in the monomer's rotational signal could be indicators of the very
first stages of where magnesium oxide turns from a gas into a proto-mineral.
Consequently, the provided spectral features may allow for detailed and
fine-tuned analysis of planet-forming regions and specifically protoplanetary
discs.


\section{Acknowledgements}

RCF wishes to acknowledge NASA grant NNX17AH15G for support of this work.
Additionally, the authors greatly acknowledge the support of the National
Science Foundation (Award Number: NSF-CHE (REU) 1359229).




\renewcommand{\baselinestretch}{1}

\newpage

\begingroup
\begin{table*}
  \centering
  \caption{The (MgO)$_2$ CcCR and MP2-F12/aug-cc-pVDZ QFF VPT2 Frequencies (in
cm$^{-1}$) with MP2/6-31+G$^*$ Double Harmonic Intensities (in km/mol in
Parentheses).}

    \begin{tabular}{l|r r r r r r r r}
\hline\hline
		
Units&CcCR & MP2-F12 & $^{24}$Mg$^{25}$Mg& $^{25}$Mg $^{25}$Mg & $^{25}$Mg $^{26}$Mg & $^{26}$Mg $^{26}$Mg & $^{24}$Mg$^{26}$Mg & Mass Average \\
$\omega_1(A_g)$    & 681.8       & 650.1 & 679.4 & 676.6 & 674.4 & 672.1 & 677.2 & 673.1 \\
$\omega_2(B_{2u})$ & 673.6 (179) & 641.1 & 671.0 & 668.2 & 665.7 & 663.2 & 668.7 & 664.5 \\
$\omega_3(B_{3u})$ & 615.0 (201) & 593.1 & 612.4 & 610.0 & 607.7 & 605.4 & 610.0 & 607.7 \\
$\omega_4(B_{1g})$ & 580.7       & 558.8 & 577.8 & 575.0 & 572.3 & 569.8 & 574.9 & 572.8 \\
$\omega_5(A_g)$    & 453.4       & 441.7 & 450.5 & 447.6 & 444.7 & 441.9 & 447.6 & 446.8 \\
$\omega_6(B_{1u})$ & 286.9 (141) & 278.6 & 285.7 & 284.5 & 283.5 & 282.4 & 284.6 & 283.7 \\
$\nu_1(A_g)$       & 672.2       & 641.2 & 669.7 & 667.2 & 665.1 & 662.8 & 667.7 & 663.7 \\
$\nu_2(B_{2u})$    & 661.8       & 630.4 & 659.3 & 656.6 & 654.3 & 651.8 & 657.1 & 653.1 \\
$\nu_3(B_{3u})$    & 605.0       & 583.2 & 602.5 & 600.2 & 597.9 & 559.4 & 563.8 & 587.4 \\
$\nu_4(B_{1g})$    & 569.4       & 548.1 & 566.5 & 563.9 & 561.3 & 558.8 & 563.8 & 561.7 \\
$\nu_5(A_g)$       & 450.3       & 439.0 & 447.4 & 444.6 & 441.7 & 438.9 & 444.6 & 443.8 \\
$\nu_6(B_{1u})$    & 284.4       & 276.7 & 283.2 & 282.1 & 281.1 & 280.0 & 282.2 & 281.4 \\
$2\nu_2$           & 1320.4      & 1257.8& 1315.3& 1197.1& 1248.7& 930.3 & 1310.9& 1225.8\\
$2\nu_3$           & 1206.7      & 1163.1& 1201.8& 1310.0& 1192.6& 1149.4& 1197.1& 1202.9\\
$2\nu_6$           & 568.2       & 553.1 & 565.9 & 563.6 & 561.5 & 559.4 & 563.8 & 562.2 \\
Zero-Point       &1639.14     &1575.90 &1631.88 &    &      &             &     1625.13 & \\
\hline
    \end{tabular}
  \label{MgO2vib}
\end{table*}
\endgroup

\begingroup

\begin{table*}
  \centering
  \caption{The CcCR QFF Zero-Point ($R_{\alpha}$ vibrationally-averaged) and
Equilibrium Structures, Rotational Constants, and Quartic and Sextic Distortion
Constants of (MgO)$_2$ and for the $^{25}$Mg and $^{26}$Mg Isotopologues.}

\resizebox{\textwidth}{!} 
{\begin{tabular}{ l l | r r r r r r r r} 
\hline\hline

&Units&CcCR & MP2-F12 & $^{24}$Mg$^{25}$Mg& $^{25}$Mg $^{25}$Mg & $^{25}$Mg $^{26}$Mg & $^{26}$Mg $^{26}$Mg & $^{24}$Mg $^{26}$Mg & Mass Average \\
 r$_0$(O$-$Mg) & & 1.859944 & 1.88708 & 1.859905 & 1.859893 & 1.859883 & 1.859846 & 1.859867 & 1.859890 \\
$\angle$(Mg$-$O$-$Mg) & & 79.999 & 79.918 & 79.998 & 79.999 & 79.999 & 79.998 & 79.997 & 79.998 \\
 $A_0$ & MHz   & 7796.0 & 7564.7 & 7796.1 & 7796.3 & 7796.4 & 7796.5 & 7796.3 & 7796.3 \\
   $B_0$ & MHz   & 7384.9 & 7186.1 & 7235.9 & 7089.4 & 6952.5 & 6817.6 & 7096.8 & 7096.2 \\
    $C_0$ & MHz   & 3787.4 & 3680.4 & 3747.9 & 3708.2 & 3670.4 & 3632.5 & 3710.2 & 3709.4 \\
 $A_1$ & MHz   & 7776.9 & 7544.8 & 7777.2 & 7777.6 & 7777.9 & 7778.3 & 7777.5 & 7777.6 \\
    $B_1$ & MHz   & 7369.1 & 7170.9 & 7220.5 & 7074.2 & 6937.7 & 6803.1 & 7081.9 & 7081.1 \\
  $C_1$ & MHz   & 3779.0 & 3671.9 & 3740.0 & 3700.2 & 3662.9 & 3624.9 & 3703.4 & 3701.7 \\
    $A_2$ & MHz   & 7796.2 & 7563.4 & 7796.5 & 7796.9 & 7797.2 & 7797.5 & 7796.8 & 7796.9 \\
  $B_2$ & MHz   & 7355.9 & 7158.8 & 7207.7 & 7061.8 & 6925.6 & 6791.3 & 7069.4 & 7068.6 \\
  $C_2$ & MHz   & 3774.3 & 3667.3 & 3734.6 & 3695.3 & 3657.5 & 3620.0 & 3696.3 & 3696.3 \\
    $A_3$ & MHz   & 7761.3 & 7530.5 & 7761.7 & 7761.9 & 7762.2 & 7762.4 & 7762.1 & 7761.9 \\
    $B_3$ & MHz   & 7403.0 & 7203.7 & 7253.2 & 7106.0 & 6968.4 & 6833.0 & 7113.3 & 7112.8 \\
  $C_3$ & MHz   & 3777.5 & 3670.5 & 3738.2 & 3698.7 & 3661.1 & 3623.4 & 3700.7 & 3699.9 \\
   $A_4$ & MHz   & 7770.3 & 7538.9 & 7770.6 & 7770.8 & 7771.0 & 7771.1 & 7770.9 & 7770.8 \\
    $B_4$ & MHz   & 7376.2 & 7177.4 & 7227.4 & 7081.2 & 6944.5 & 6809.9 & 7088.5 & 7088.0 \\
 $C_4$ & MHz   & 3773.7 & 3666.8 & 3734.3 & 3694.7 & 3657.1 & 3619.3 & 3696.7 & 3696.0 \\
     $A_5$ & MHz   & 7811.9 & 7580.8 & 7811.8 & 7811.8 & 7811.7 & 7811.6 & 7811.7 & 7811.8 \\
     $B_5$ & MHz   & 7384.9 & 7185.0 & 7236.0 & 7089.5 & 6952.6 & 6817.8 & 7097.0 & 7096.3 \\
     $C_5$ & MHz   & 3784.9 & 3677.8 & 3745.3 & 3705.7 & 3667.9 & 3630.0 & 3707.7 & 3706.9 \\
     $A_6$ & MHz   & 7780.4 & 7548.8 & 7780.4 & 7780.3 & 7811.7 & 7811.6 & 7780.3 & 7790.8 \\
     $B_6$ & MHz   & 7361.2 & 7162.3 & 7213.2 & 7067.5 & 6952.6 & 6817.8 & 7074.9 & 7081.2 \\
     $C_6$ & MHz   & 3791.0 & 3683.7 & 3751.4 & 3711.6 & 3667.9 & 3630.0 & 3713.6 & 3710.9 \\
\hline
     r$_e$(O$-$Mg) & \AA   & 1.853962 & 1.880792 & 1.853962 & 1.853962 & 1.853962 & 1.853962 & 1.853962 & 1.853962 \\
     $\angle$(Mg$-$O$-$Mg) & $^{\circ}$ & 80.026 & 79.951 & 80.026 & 80.026 & 80.026 & 80.026 & 80.026 & 80.026 \\
     $A_e$ & MHz   & 7835.43 & 7605.11 & 7835.43 & 7835.43 & 7835.43 & 7835.43 & 7835.43 & 7835.43 \\
     $B_e$ & MHz   & 7414.35 & 7215.57 & 7264.66 & 7117.37 & 6979.82 & 6844.33 & 7124.89 & 7124.24 \\
     $C_e$ & MHz   & 3809.54 & 3702.61 & 3769.63 & 3729.58 & 3691.46 & 3653.21 & 3731.64 & 3730.84 \\
     $D_J$ & kHz & 3.296 & 3.1755 & 3.1542 & 3.0201 & 2.9003 & 2.7862 & 3.028 & 3.0308 \\
     $D_{JK}$ & kHz   & -3.1362 & -2.6168 & -2.8306 & -2.559 & -2.3299 & 2.7862 & -2.5735 & -1.7738 \\
     $D_K$ & kHz   & 6.6563 & 6.2184 & 6.4924 & 6.355 & 6.2456 & 6.154 & 6.3615 & 6.3775 \\
     $d_1$ & kHz   & -1.6033 & -1.5492 & -1.5363 & -1.4714 & -1.4123 & -1.355 & -1.4753 & -1.4756 \\
     $d_2$ & kHz   & -0.226 & -0.2331 & -0.2253 & -0.2229 & -0.2195 & -1.355 & -0.2231 & -0.4120 \\
     $H_J$ & Hz    & 0.00664 & 0.0062 & 0.00604 & 0.00549 & 0.00503 & 0.00461 & 0.00552 & 0.00556 \\
     $H_{JK}$ & Hz    & -0.0302 & -0.02713 & -0.02683 & -0.02386 & -0.02142 & -0.01925 & -0.02407 & -0.02427 \\
     $H_{KJ}$ & Hz    & 0.00459 & -0.00095 & 0.00122 & -0.00165 & -0.00387 & -0.00577 & -0.00141 & -0.00115 \\
     $H_K$ & Hz    & 0.03331 & 0.0362 & 0.03392 & 0.03437 & 0.03461 & 0.03476 & 0.0343 & 0.0342 \\
     $h_1$ & Hz    & 0.003 & 0.00284 & 0.00282 & 0.00266 & 0.00251 & 0.00236 & 0.00267 & 0.00267 \\
     $h_2$ & Hz    & -0.00013 & -0.00011 & -0.00001 & 0.00008 & 0.00015 & 0.00021 & 0.00008 & 0.00006 \\
     $h_3$ & Hz    & 0.0002 & 0.00016 & 0.00018 & 0.00017 & 0.00016 & 0.00015 & 0.00017 & 0.00017 \\
\hline
     \end{tabular}}
  \label{MgO2geom}
\end{table*}

\endgroup

\newpage

\makeatletter
\setlength{\@fptop}{0pt}
\makeatother

\begingroup
\begin{table*}
  \centering
  \caption{The MP2-F12/aug-cc-pVDZ QFF VPT2 Magnesium Oxide Timer
Frequencies (in cm$^{-1}$) with MP2/6-31+G$^*$ Double Harmonic Intensities (in
km/mol in Parentheses).}
\scriptsize
      \begin{tabular}{l|r r r r r r r r}
\hline\hline
		
Units& $^{24}$Mg$^{24}$Mg$^{24}$Mg & $^{24}$Mg$^{24}$Mg$^{25}$Mg& $^{24}$Mg $^{25}$Mg $^{25}$Mg & $^{24}$Mg $^{26}$Mg $^{26}$Mg & $^{24}$Mg $^{24}$Mg $^{26}$Mg & $^{26}$Mg $^{24}$Mg $^{25}$Mg & Mass Average \\

$\omega_1(A_2')$  & 782.3       & 780.2 & 777.8 & 774.6 & 778.9 & 776.3 & 778.3 \\
$\omega_2(E')$    & 757.6 (321) & 757.4 & 754.6 & 751.0 & 757.1 & 753.7 & 755.2 \\
                  &             & 753.0 & 751.2 & 745.4 & 748.3 & 747.4 & 749.1 \\
$\omega_3(E')$    & 600.9 (1)   & 599.7 & 598.3 & 595.9 & 598.6 & 597.2 & 598.4 \\
                  &             & 599.4 & 598.1 & 595.5 & 598.0 & 596.7 & 597.5 \\
$\omega_4(A_1')$  & 513.3       & 512.6 & 511.8 & 510.5 & 511.9 & 511.1 & 511.9 \\
$\omega_5(A_1')$  & 355.0       & 353.1 & 351.2 & 347.7 & 351.3 & 349.5 & 351.3 \\
$\omega_6(A_2'')$ & 269.2 (220) & 268.4 & 267.7 & 266.4 & 267.8 & 267.1 & 267.8 \\
$\omega_7(E')$    & 188.4 (82)  & 188.1 & 187.2 & 186.1 & 187.8 & 186.8 & 187.4 \\
                  &             & 187.0 & 186.2 & 184.2 & 185.7 & 185.1 & 185.6 \\
$\omega_8(E'')$   & 165.9       & 165.9 & 165.3 & 164.8 & 165.9 & 165.2 & 165.5 \\
                  &             & 164.8 & 164.3 & 162.9 & 163.8 & 163.5 & 164.2 \\
$\nu_1(A_2')$     & 767.8       & 765.0 & 763.3 & 759.5 & 764.6 & 761.8 & 763.7 \\
$\nu_2(E')$       & 751.2       & 748.4 & 747.4 & 744.1 & 751.9 & 748.4 & 748.6 \\
                  &             & 749.6 & 746.0 & 740.2 & 740.3 & 740.2 & 743.2 \\
$\nu_3(E')$       & 597.8       & 594.5 & 594.2 & 591.8 & 596.6 & 594.6 & 594.9 \\
                  &             & 598.5 & 596.1 & 593.4 & 593.9 & 593.2 & 595.0 \\
$\nu_4(A_1')$     & 514.1       & 513.0 & 512.1 & 510.3 & 512.3 & 511.3 & 512.2 \\
$\nu_5(A_1')$     & 374.9       & 372.3 & 370.7 & 366.7 & 371.7 & 369.1 & 370.9 \\
$\nu_6(A_2'')$    & 314.2       & 312.9 & 312.1 & 310.1 & 312.8 & 311.4 & 312.2 \\
$\nu_7(E')$       & 216.4       & 204.7 & 209.1 & 208.1 & 221.2 & 214.7 & 212.4 \\
                  &             & 226.0 & 219.5 & 216.5 & 207.4 & 211.9 & 216.3 \\
$\nu_8(E'')$      & 194.3       & 197.5 & 195.3 & 194.5 & 192.6 & 193.4 & 194.6 \\
                  &             & 189.8 & 190.6 & 188.8 & 193.3 & 191.1 & 190.7 \\
Zero-Point        & 2718.69     &2710.1 &2702.1 &2686.8 &2703.2 &2694.7 &2702.6 \\
\hline
     \end{tabular}
  \label{MgO3vib}
\end{table*}

\endgroup

\begingroup
\begin{table*}
  \centering
  \caption{The (MgO)$_3$ MP2-F12/aug-cc-pVDZ QFF Zero-Point ($R_{\alpha}$
vibrationally-averaged) and Equilibrium Structures, Rotational Constants, and
Quartic and Sextic Distortion Constants Including Some of the $^{25}$Mg
and $^{26}$Mg Isotopologues.}
\tiny
\begin{tabular}{ l l | r r r r r r r } 
\hline\hline

&Units& $^{24}$Mg$^{24}$Mg$^{24}$Mg & $^{24}$Mg$^{24}$Mg$^{25}$Mg& $^{24}$Mg $^{25}$Mg $^{25}$Mg & $^{24}$Mg $^{26}$Mg $^{26}$Mg & $^{24}$Mg $^{24}$Mg $^{26}$Mg & $^{26}$Mg $^{24}$Mg $^{25}$Mg & Mass Average \\

       r$_0$(O$-$Mg) &   \AA  & 1.848368 & 1.848348 & 1.84834 & 1.848318 & 1.848356 & 1.848336 & 1.848344 \\
     {$\angle$(Mg$-$O$-$Mg)} &    $^{\circ}$ & 73.643 & 73.643 & 73.642 & 73.643 & 73.644 & 73.644 & 73.643 \\
       $A_0$ & MHz   & 2579.2 & 2572.0 & 2559.7 & 2526.7 & 2563.0 & 2538.0 & 2556.4 \\
       $B_0$ & MHz   & 2561.2 & 2531.2 & 2506.5 & 2469.0 & 2505.2 & 2493.1 & 2511.0 \\
       $C_0$ & MHz   & 1284.0 & 1274.7 & 1265.4 & 1247.7 & 1265.8 & 1256.6 & 1265.7 \\ \hline
       r$_e$(O$-$Mg) & \AA   & 1.844533 & 1.844533 & 1.844533 & 1.844533 & 1.844533 & 1.844533 & 1.844533 \\
       $\angle$(Mg$-$O$-$Mg) & $^{\circ}$ & 73.622 & 73.622 & 73.622 & 73.622 & 73.622 & 73.622 & 73.622 \\
       $A_e$ & MHz   & 2586.77 & 2580.08 & 2567.29 & 2534.96 & 2571.21 & 2546.28 & 2564.43 \\
       $B_e$ & MHz   & 2569.74 & 2539.14 & 2514.7 & 2476.11 & 2512.79 & 2500.35 & 2518.81 \\
       $C_e$ & MHz   & 1289.11 & 1279.72 & 1270.36 & 1252.59 & 1270.83 & 1261.55 & 1270.69 \\
       $D_J$ & kHz & 0.3046 & 0.3101 & 0.2947 & 0.2898 & 0.3155 & 0.3009 & 0.3026 \\
       $D_{JK}$ & kHz   & 0.9528 & 0.8783 & 0.942 & 0.9253 & 0.8091 & 0.885 & 0.899 \\
       $D_K$ & kHz   & -0.2212 & -0.1596 & -0.2237 & -0.2365 & -0.1042 & -0.1919 & -0.1895 \\
       $d_1$ & kHz   & -0.2444 & -0.2424 & -0.2357 & -0.2314 & -0.241 & -0.2368 & -0.2386 \\
       $d_2$ & kHz   & -0.1123 & -0.1073 & -0.1081 & -0.1059 & -0.103 & -0.1059 & -0.1071 \\
       $H_J$ & Hz    & -0.0001 & 0.00014 & -0.00063 & -0.00062 & 0.00053 & -0.00011 & -0.00013 \\
       $H_{JK}$ & Hz    & 0.0026 & -0.00125 & 0.0127 & 0.01256 & -0.00851 & 0.00301 & 0.00352 \\
       $H_{KJ}$ & Hz    & -0.00142 & 0.00819 & -0.02898 & -0.02866 & 0.0273 & -0.00313 & -0.00445 \\
       $H_K$ & Hz    & 0.00024 & -0.00599 & 0.01877 & 0.01848 & -0.01865 & 0.00148 & 0.00239 \\
       $h_1$ & Hz    & 0.00022 & 0.00022 & 0.00025 & 0.00024 & 0.00019 & 0.00022 & 0.00022 \\
       $h_2$ & Hz    & 0.00037 & 0.00031 & 0.00047 & 0.00046 & 0.0002 & 0.00035 & 0.00036 \\
       $h_3$ & Hz    & 0.0001 & 0.00016 & -0.00009 & -0.00009 & 0.00027 & 0.00008 & 0.00007 \\
\hline
       \end{tabular}
  \label{MgO3geom}
\end{table*}
\endgroup

\end{document}